 \documentclass[final,3p,times]{elsarticle}
\usepackage{graphicx}

\usepackage{amssymb}
\usepackage{amsmath}
\journal{Proceedings of the Combustion Institute}

\begin{document}
%\small
%\baselineskip 10pt
\begin{frontmatter}

%% Title, authors and addresses

%% use the tnoteref command within \title for footnotes;
%% use the tnotetext command for the associated footnote;
%% use the fnref command within \author or \address for footnotes;
%% use the fntext command for the associated footnote;
%% use the corref command within \author for corresponding author footnotes;
%% use the cortext command for the associated footnote;
%% use the ead command for the email address,
%% and the form \ead[url] for the home page:
%%
%% \title{Title\tnoteref{label1}}
%% \tnotetext[label1]{}
%% \author{Name\corref{cor1}\fnref{label2}}
%% \ead{email address}
%% \ead[url]{home page}
%% \fntext[label2]{}
%% \cortext[cor1]{}
%% \address{Address\fnref{label3}}
%% \fntext[label3]{}
\title{Detonation re-initiation mechanism following the Mach reflection of a quenched detonation}
%% use optional labels to link authors explicitly to addresses:
%% \author[label1,label2]{<author name>}
%% \address[label1]{<address>}
%% \address[label2]{<address>}

\author{R. Bhattacharjee, S.SM. Lau-Chapdelaine, G. Maines, L. Maley, M.I. Radulescu}

\address{Department of Mechanical Engineering, University of Ottawa, Ottawa Canada K1N6N5}

\begin{abstract}
This experimental study addresses the re-initiation mechanism of detonation waves following the Mach reflection of a shock-flame complex.  The detonation diffraction around a cylinder is used to reproducibly generate the shock-flame complex of interest. The experiments are performed in methane-oxygen.  We use a novel experimental technique of coupling a two-in-line-spark flash system with a double-frame camera in order to obtain microsecond time resolution permitting accurate schlieren velocimetry. The first series of experiments compares the non-reactive sequence of shock reflections with the reflection over a rough wall under identical conditions.  It was found that the hot reaction products generated along the rough wall are entrained by the wall jet into a large vortex structure behind the Mach stem. The second series of experiments performed in more sensitive mixtures addressed the sequence of events leading to the detonation establishment along the Mach and transverse waves. Following ignition and jet entrainment, a detonation first appears along the Mach stem while the transverse wave remains non-reactive.  The structure of the unburned tongue however indicates local instabilities and hot spot formation, leading to the rapid reaction of this gas.  Numerical simulations are also reported, confirming the sequence of ignition events obtained experimentally.   
\end{abstract}

\begin{keyword}
reactive Mach reflection \sep time resolved schlieren photography \sep gaseous detonations \sep Mach jet \sep amplification mechanism \sep Kelvin-Helmholtz instability

%% MSC codes here, in the form: \MSC code \sep code
%% or \MSC[2008] code \sep code (2000 is the default)

\end{keyword}

\end{frontmatter}
%%
%% Start line numbering here if you want
%%
% \linenumbers
%% main tex
\section{Introduction}
\begin{figure}
\centering
\includegraphics[width=0.5\columnwidth]{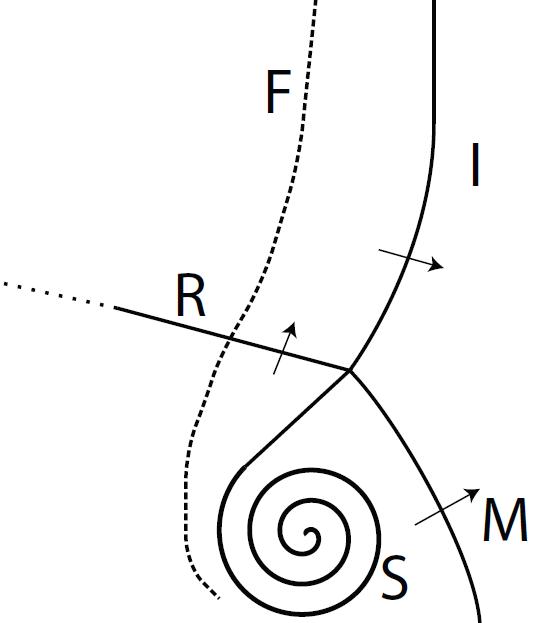}
\caption{Schematic of Mach reflection of a decoupled shock-flame complex: (M) Mach shock, (I) incident shock, (F) flame, (R) reflected or transverse wave, (S) slip-line.} 
\label{fig:fig1}
\end{figure}

\addvspace{10pt}
The present study addresses the re-initiation mechanism occurring when a decoupled shock-flame complex undergoes a Mach reflection process, as illustrated in Figure \ref{fig:fig1}.  This problem arises within the detonation wave structure, where new overdriven detonations are periodically reformed following irregular Mach reflections \cite{Lee2008, Shepherd2009}. To date, five distinct mechanisms have been suggested to play an important role in igniting substantial amount of gases during this transient Mach reflection process \cite{Subbotin1975b, Austinetal2005, Radulescuetal2005, Radulescuetal2007,  Kessleretal2011, Mahmoudi&Mazaheri2011, Gamezoetal2000}: 
\begin{enumerate}
\item the much higher temperature behind the Mach stem leads to a significant reduction of the ignition delays and rapid reaction of the gases,
\item Kelvin-Helmholtz instability along the slip-line emanating from the triple points promotes turbulent mixing between reacted and un-reacted gases, leading to increased burning rates,
\item Richtmyer-Meshkov instability occurring when transverse shocks interact with the flame promote turbulent mixing and enhance the burning rates,
\item rapid reactions behind the transverse shocks lead to transverse detonations, which rapidly consume the gas, and,
\item the strong jet formation during the shock reflection process enhances the mixing behind the Mach stem and leads to the reaction of substantial amount of gases.  
\end{enumerate}

While the first mechanism is commonly accepted, evidence in support for the other four mechanisms remains inconclusive.  This is principally due to the difficulty in experimentally probing the flowfield within the detonation structure, particularly for highly unstable detonations. While current numerical studies provide the most telling evidence, they themselves present limitations due to simplifying assumptions about the chemical kinetics, molecular transport and resolution.  The present study is an experimental attempt to isolate the key mechanisms that contribute to the acceleration of the reaction rates during Mach shock reflections in reactive gases.

The Mach reflection occurring within the detonation wave structure is also very similar to the reflection of a shock-flame complex, following a detonation diffracting around an obstacle or a backward facing step.  This type of configuration presents results on a larger scale than cellular dynamics and therefore allows researchers to make detailed observations.  Teodorczyk, Lee and Knystautas  studied the wall-reflection of a detonation diffracted around a baffle using high speed schlieren visualization \cite{Teodorczyketal1988, Teodorczyketal1990}. From their experiments, they clearly correlated the re-ignition of the gas with the region behind the Mach stem.  These authors have also pointed out that mixing of hot reacted gases with colder un-reacted gases along the slip-line emanating from the triple points may also have a strong influence.  Ohyagi et al. \cite{Ohyagietal2002} and Obara et al. \cite{Obaraetal2008} also studied the diffraction and re-initiation events that take place behind backward facing steps. Their high-speed schlieren images and soot foils clearly indicated that re-initiation takes place behind the Mach stem.  They also provided evidence that the re-initiation event is accompanied by transverse detonations.  Such transverse detonations have also been inferred from soot foils \cite{Sorinetal2009} and open shutter photographs \cite{Radulescu&Maxwell2011}.  

Open shutter photography by Radulescu and Maxwell has shown evidence of intense chemi-luminescence near triple points prior to the re-initiation  \cite{Radulescu&Maxwell2011}.  From their analysis,  as transverse detonations were ruled out, the evidence pointed towards the possibility of turbulent mixing.  The experiments, however, could not identify the true mechanism responsible for the intense chemi-luminescence observed.  

A mechanism that has been suggested to play a role in shock reflections leading to detonation re-initiation is the jet formed in irregular shock reflections \cite{Subbotin1975, Radulescuetal2009, Mach&Radulescu2011, Collela&Glaz1984}. The forward jetting of the shear layer in Mach reflections can potentially re-inject combustion products of the trailing reaction zone into the un-reacted shocked gas behind the Mach stem, as proposed by Sorin et al. \cite{Sorinetal2009}. Indeed, very recent numerical simulations of cellular detonations performed by Mahmoudi and Mazaheri \cite{Mahmoudi&Mazaheri2011} have shown that this jet plays a critical role in increasing the number of hot spots (isolated regions of intense chemical reactivity) in the unburnt gases behind the Mach stem.

The technique of using large scale detonation quenching by an obstacle and subsequent re-ignition via shock reflections offers the opportunity to study the re-ignition by Mach shock reflections on larger scales than cellular detonations.  This is the strategy employed in the present study: we use the detonation diffraction around a half-cylinder to provide a quenched detonation, which is then followed by a wall reflection generating the Mach reflection of interest. Using high spatial and temporal resolution visualization, we aim to isolate the important mechanisms controlling the re-initiation of the decoupled shock-flame complex into an overdriven detonation.

%%%%%%%%%%%%%%%%%%%%%%%%%%%%%%%%%%%%%%%%%%%%%%%%%%%%%%%%%%%

\section{Experimental technique}
\begin{figure}
\centering
 \includegraphics[width=\columnwidth]{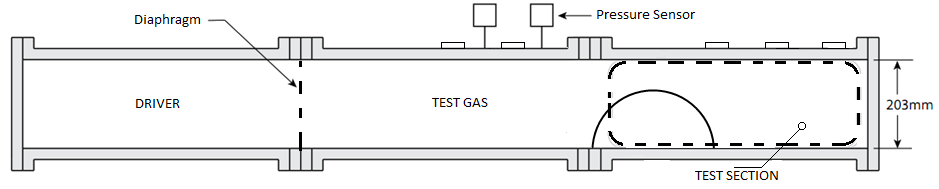}
\caption{Schematic of the shock tube apparatus}
\label{fig:fig2}
\end{figure}

The experiments were carried out in a shock tube illustrated schematically in Figure \ref{fig:fig2}. The channel has a rectangular cross-section of 19 mm by 203 mm.  Its high aspect ratio permits to establish essentially two-dimensional flow-fields.  The shock tube is approximately 3.4 m long and consists of three aluminium sections of equal length. The first section served as a reactive driver, while the second and third sections contained the test gas.  The driver gas (acetylene-oxygen) was separated from the test gas by a thin plastic diaphragm.  The third section of the channel has an unobstructed field of view, which served as the viewing area for the photographic setup.  The obstacle used for the experiments is a polyoxymethylene thermoplastic (Delrin) half cylinder with a radius of 152 mm. 

All experiments were performed in a test mixture of stoichiometric methane-oxygen.  Gases were prepared before an experiment in a separate vessel by the method of partial pressures.  Varying the initial pressure of the test mixture permitted us to control the reactivity of the mixture \cite{Radulescu&Maxwell2011}.  Before every experiment, the shock tube was first cleared of any debris from previous experiments, then evacuated to approximately 75 Pa before injecting our test gas. Pressure transducers positioned upstream of the obstacle permitted us to monitor whether a self-sustained detonation was initiated before its diffraction over the obstacle.  

A \textit{Z}-configuration schlieren setup was used, using a two-spark light source (PalFlash 501).  The flash system delivers two separately timed flashes along the same optical axis, each flash lasting approximately 250 ns.  In all experiments a 11 $\mu{s}$ delay was used between the successive frames, recorded with a double frame camera (PCO-1600) with 1600 by 1200 pixel resolution.  The double frame capability permitted us to perform velocimetry of the shock evolution between the two sequential images.  Such velocity calculations were carried out along the top wall and bottom wall of the shock tube.  The errors associated with computing the shock speed from our image analysis was estimated to be less than 8$\%$.  

In all photographs, a vertical schlieren knife edge was used, which permitted us to visualize horizontal density gradients.  For reference, positive horizontal density gradients are lighter while negative density gradients are darker, as for example right facing shock waves in the photographs.

%%%%%%%%%%%%%%%%%%%%%%%%%%%%%%%%%%%%%%%%%%%%%%%%%%%%%%%%%%%%

\section{The non-reactive shock reflection details}
\begin{figure}
\centering
\includegraphics[width=\columnwidth]{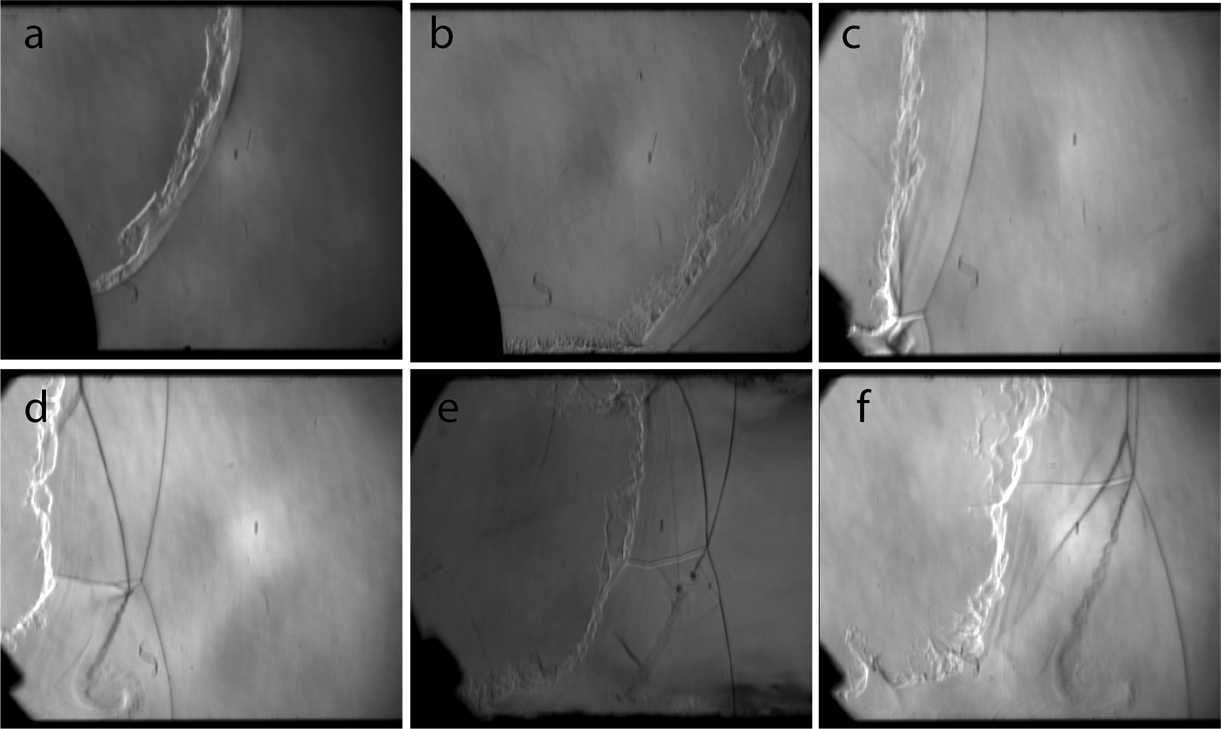}
\caption{Schlieren photographs illustrating the detonation quenching around the cylinder and the subsequent shock reflections at $p_0$=5.5 kPa.}
\label{fig:55kpa}
\end{figure}

At an initial pressure of $p_0$= 5.5 kPa, the diffraction of the detonation around the cylinder resulted in a decoupled shock-flame complex.  Owing to the experimental reproducibility, Figure \ref{fig:55kpa} shows the evolution of the flowfield constructed from separate experiments.  Figure \ref{fig:55kpa}a shows the detonation diffracting around the cylinder, with the decoupled flame behind the curved shock.  Figure \ref{fig:55kpa}b shows the initial regular reflection of the incident shock, which becomes a Mach reflection in the subsequent frames. From our image analysis, at the onset of an irregular reflection as shown in Figure \ref{fig:55kpa}c (estimated to be approximately 300 mm from the center of the obstacle), typical strengths of the Mach stem and incident wave are $M=4.1$ and $M=3.3$ respectively. We also observed that the shock velocity decreases gradually as the non-reactive shock structure travels downstream.
	
Figures \ref{fig:55kpa}c-d clearly show the familiar double Mach reflection configuration.  Also apparent in Figures \ref{fig:55kpa}c-d is the reflected transverse wave, which travels faster through the combustion products, owing to the much higher sound speed in the products.  In Figure \ref{fig:55kpa}d, this reflected wave has reached the top wall, reflected and formed an irregular reflection; the slip-line is not visible due to the weak density gradients across it. 

This reflected wave gives rise to a new shock wave in the non-reacted layer of gas.  In Figure \ref{fig:55kpa}d, this inner shock is about to reach the main triple point.  In Figure \ref{fig:55kpa}e, the reflected wave meets with the main triple point, which has now become a \emph{quadruple point};  the \emph{quadruple point} is the confluence of four shocks and a slip-line.  We find that this \emph{quadruple point} remains stable for a short transient, as illustrated in Figures \ref{fig:55kpa}e-f .  To the best of our knowledge, this is the first observation of such a flow structure. 

The evolution of the flow-field does show indications of hydrodynamic instabilities.  Figures \ref{fig:55kpa}d-f show that the termination of the slip-line is a wall jet \cite{Mach&Radulescu2011}, indicated by the curl-up of the slip-line behind the Mach shock. Other than the prominent wall jet, Figures \ref{fig:55kpa}d-f clearly show the characteristic volutes of Kelvin-Helmholtz instability along the slip line.  However, Figures \ref{fig:55kpa}c-d shows that the refraction of the reflected wave at the flame surface does not appear to entrain significant turbulence, as the flame surface does not develop any visible corrugations or visible hot spot activity.   

%%%%%%%%%%%%%%%%%%%%%%%%%%%%%%%%%%%%%%%%%%%%%%%%%%%%%%%%%%%%

\section{Rough-wall induced ignition}
\begin{figure}
\centering
\includegraphics[width=\columnwidth]{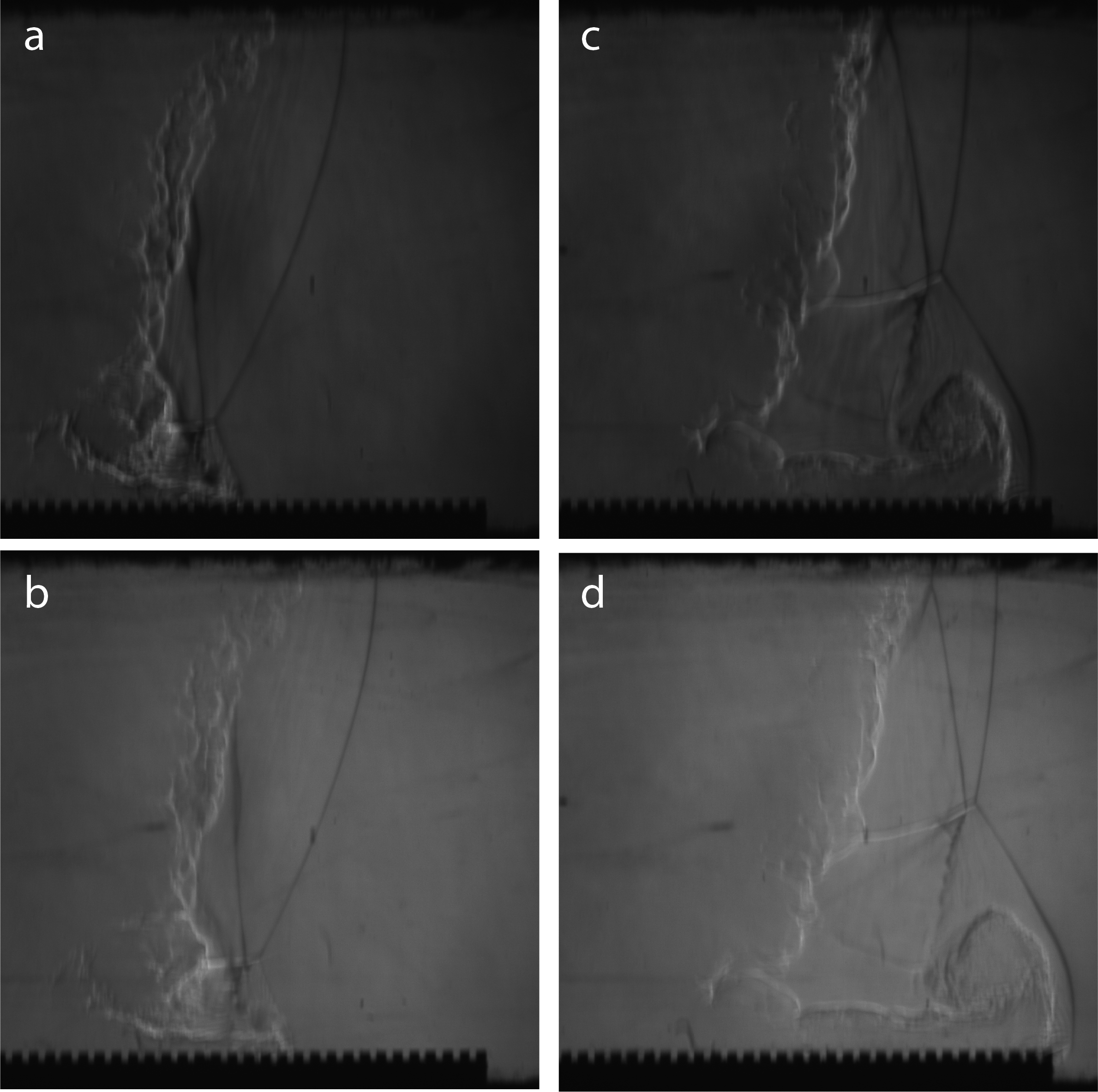}
\caption{Re-ignition and vortex entrainment on a rough plate at $p_0$=5.5 kPa.}
\label{fig:rough}
\end{figure}

The non-reactive flowfield above served as a basis for studying the ignition events and modifications due to energy release. The second series of experiments, using the same initial conditions as above, was carried out by lining the bottom wall with a rough plate. This rough plate was machined, using the same material as the half-cylinder obstacle, to a height of 10 mm with regularly spaced square grooves 3.2 mm deep and spaced by 3.2 mm.  The intention was to artificially create localized hot spots in the grooves, such that a source of combustion products appears near the bottom wall.  Figure \ref{fig:rough} shows two image pairs obtained from two separate experiments.  

In Figures \ref{fig:rough}a and \ref{fig:rough}b, the layer of burned gas appears at the surface of the rough plate and extends approximately 5 mm above the wall.  The volume occupied by this reacted gas is consistent with our estimate of the 3-fold volumetric expansion of the gas that was initially confined inside the grooves, once it has ignited. This was estimated using the Gordon-McBride equilibrium code \cite{Gordon&McBride} by assuming the gas equilibrates to the same post Mach stem pressure and maintains a constant enthalpy. 

The strong modification added by the surface ignition events to the flow-field observed above in Figure \ref{fig:55kpa} is the large vortex observed behind the Mach shock.  Figure \ref{fig:rough}c and Figure \ref{fig:rough}d shows that the filamentary structure of this vortex roll rapidly disappears.  For reference, the strength of the Mach stem in Figure \ref{fig:rough}c was evaluated to be $M=4.1$, for which chemical induction calculations using the GRI-3 mechanism and the Cantera package \cite{Goodwin} indicate an ignition delay of approximately 200 ms.  From Figure \ref{fig:rough}, the delay between the first and second image pair is approximately 50 $\mu{s}$.  The presence of chemical reactions very close to the Mach stem is thus completely incompatible with a shock-induced ignition scenario.  Instead, the vortex observed is much more likely to be associated with the wall jet observed above (Figure \ref{fig:55kpa}) for the smooth wall experiments, and in numerical simulations (see below).  

%%%%%%%%%%%%%%%%%%%%%%%%%%%%%%%%%%%%%%%%%%%%%%%%%%%%%%%%%%%%

\section{Reflections with self-ignition}
\begin{figure}
\centering
\includegraphics[width=\columnwidth]{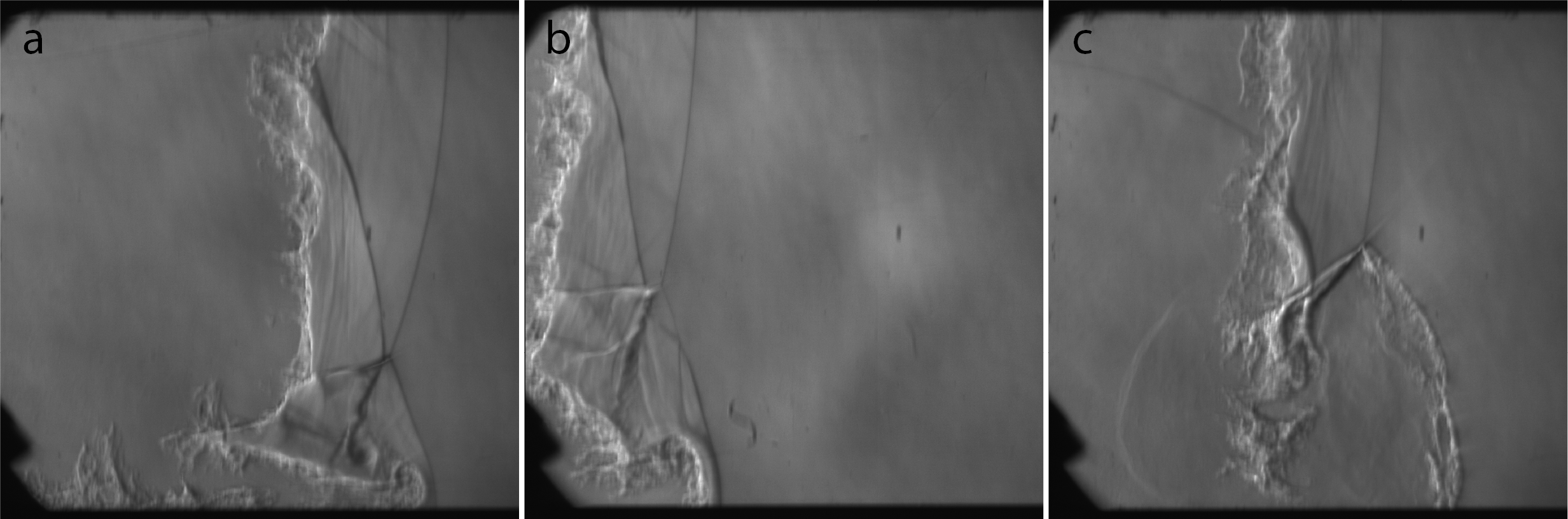}
\caption{Re-ignition and vortex entrainment at $p_0$=10 kPa.}
\label{fig:10kpa}
\end{figure}

At higher initial pressures, self-ignition was observed behind the Mach stem in smooth wall experiments.  Figure \ref{fig:10kpa} shows examples of ignition events observed at $p_0=10$ kPa. Note that a higher pressure reduces the ignition delays, which are inversely proportional to pressure, all other conditions kept constant.  At $p_0=10$ kPa, repeat experiments did not reproducibly give exactly the same flow-field, although a general sequence was established. 

Figure \ref{fig:10kpa}a show that ignition spots first appeared near the bottom wall, presumably from the Mach shock compression. This was verified through our numerical simulations, presented below. Figures \ref{fig:10kpa}a-b show the characteristic wall jet entraining product gases behind the Mach stem.  Figure \ref{fig:10kpa}c illustrates a detonation wave with the distinctive cellular structure along the Mach stem surface. The tongue of un-reacted gas behind the reflected shock also shows partial reaction, as it develops holes.  Note also in Figure \ref{fig:10kpa}c the rear facing pressure wave;  our interpretation of this event is the rapid combustion of the gases, either behind the Mach stem on in the un-reacted tongue.  

\begin{figure}
\centering
\includegraphics[width=\columnwidth]{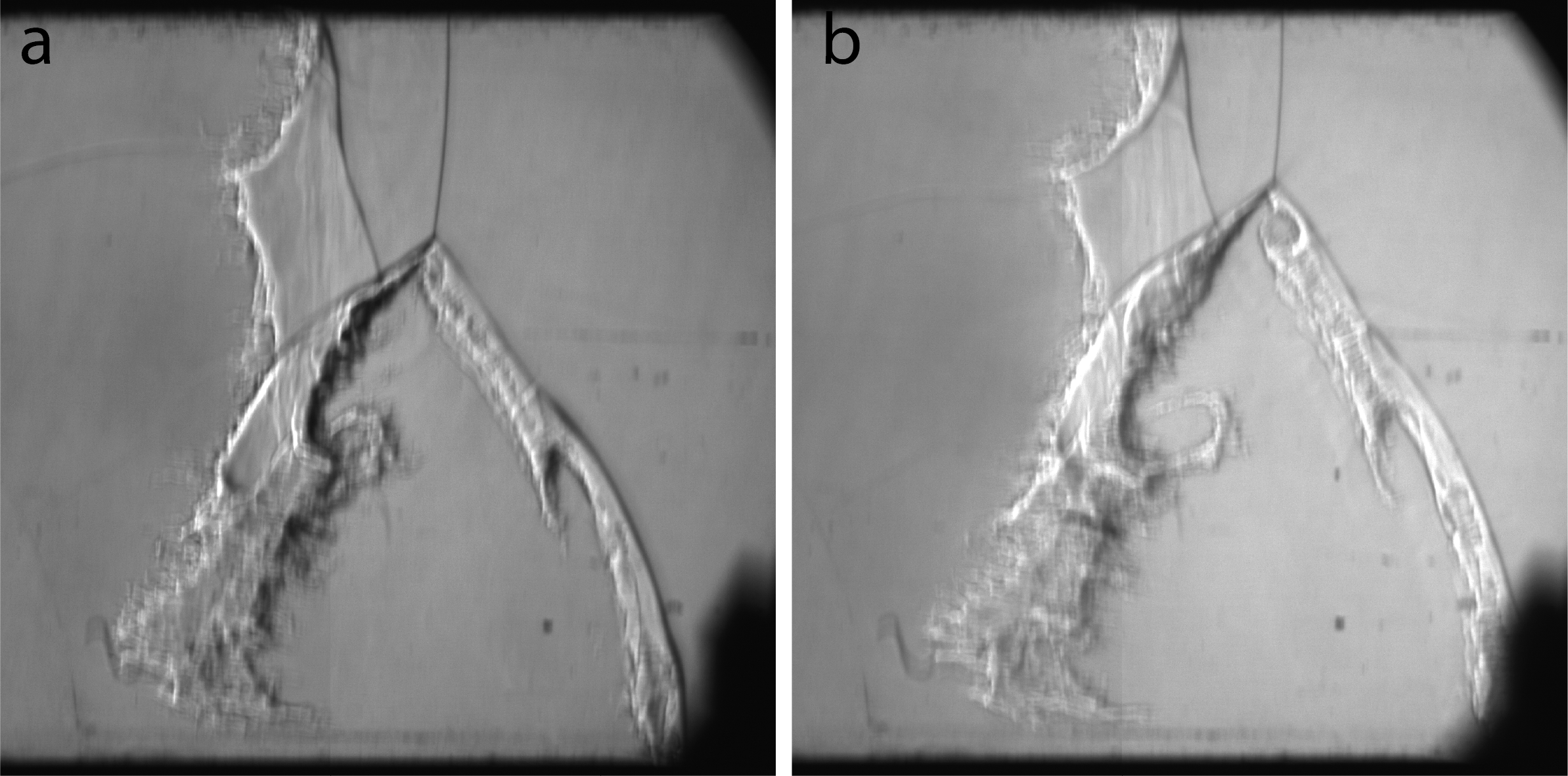}
\caption{Sequential frames at 11 $\mu{s}$ interval illustrating the local hot spot evolution at $p_0$=10 kPa.}
\label{fig:detail10kpa}
\end{figure}

Two sequential frames further illustrating the complex flow occurring during this transient are shown in Figure \ref{fig:detail10kpa}.  The reflected wave also takes on a complex pattern of shock refraction, very similar to photographs presented by Radulescu et al. \cite{Radulescuetal2007}. In these two photographs, the Mach stem has not yet established a detonation, although localized coupling can be observed. The Mach stem strength here was estimated to be approximately $M=5.0$.  Localized hot-spots can also be observed along the reaction layer behind the Mach shock, which grow between the successive frames (for example, near the triple point). Figure \ref{fig:detail10kpa} also shows evidence of Kelvin-Helmholtz instability along the slip-line.  Near the center of the photograph, we can clearly observe the rapid thinning-out of the vortex like structure of reacted gas.  The interior of the un-reacted tongue also develops holes, presumably though local reactions.

\begin{figure}
\centering
\includegraphics[width=\columnwidth]{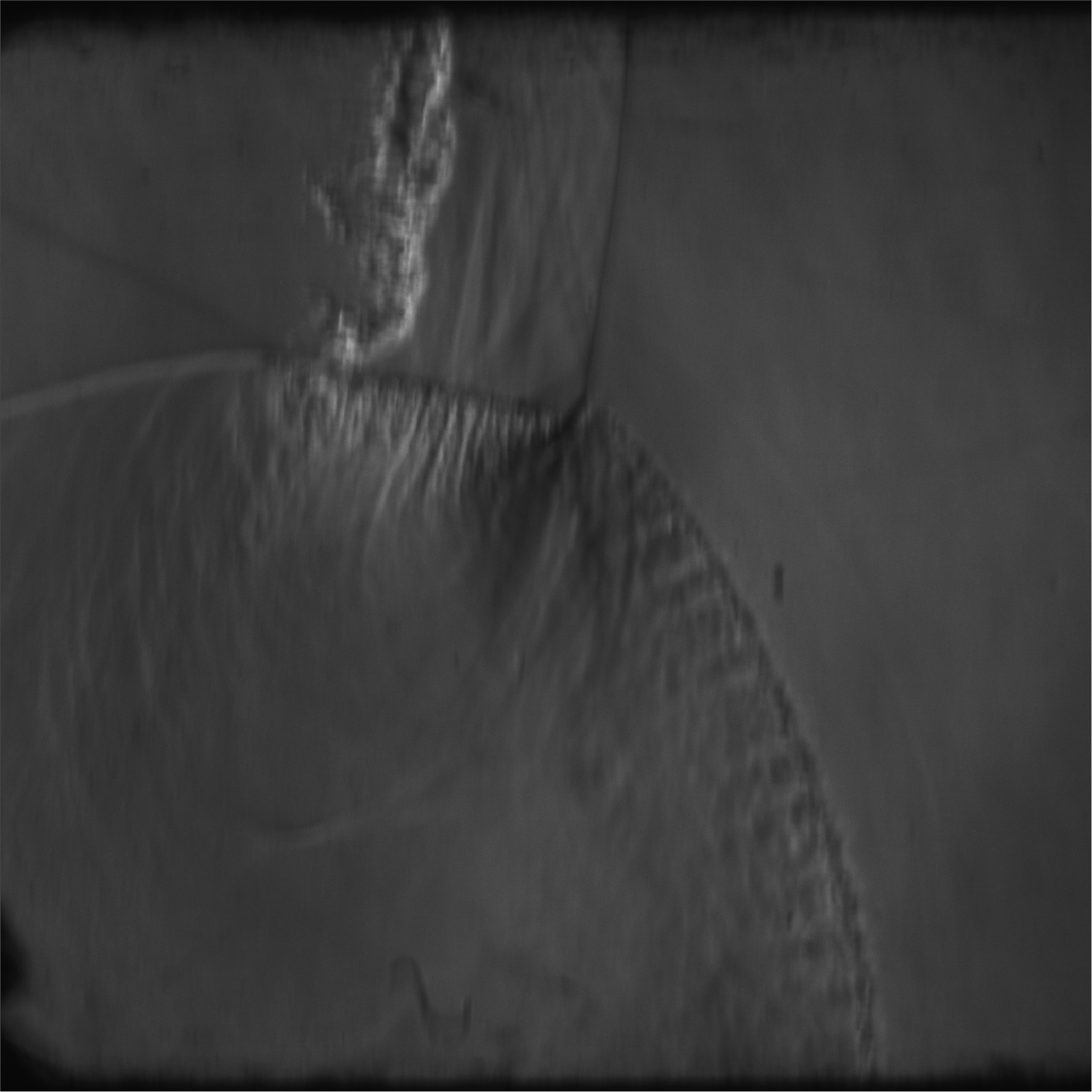}
\caption{Transverse detonation establishment at $p_0$=12 kPa}
\label{fig:12kpa}
\end{figure}

Further increasing the pressure to $p_0 =12$ kPa systematically yielded detonation wave re-establishment along both the Mach shock and the transverse wave; Figure \ref{fig:12kpa} is a very reproducible example.  In this photograph, we see that all the gas behind the transverse wave is reacted.  The transverse wave also displays the characteristic fine cellular structure expected of transverse detonations, as predicted by Gamezo et al. \cite{Gamezoetal2000} and observed for single head spin detonations (see for example Ref. \cite{Lee2008}).  Our image analysis revealed that these transverse detonations propagate at approximately $90\%$ of the Chapman-Jouguet value, calculated by taking the non-reacted gas behind the $M=$3.0 incident shock as the initial condition.  This confirmed that indeed these reflected waves are detonation waves.  Note also that from Figure \ref{fig:12kpa} the transverse detonation is not directly attached to the point where the incident wave and Mach detonation stem meet.  This observation agrees with the results reported by Gamezo et al. \cite{Gamezoetal2000} in their numerical simulations. 

Also noticeable in Figure \ref{fig:12kpa} are two non-reactive shock waves, extending towards the left from the transverse detonation, into the reacted gas behind the flame structure. The shock wave on the top (black) is simply an extension of the transverse detonation, while the shock wave below it (white) may have originated from the rapid combustion of the tongue of unburnt gas shown in Figure \ref{fig:detail10kpa}.  Indeed, as pointed earlier in Figure \ref{fig:10kpa}c, we do see shock waves generated most likely by rapid combustion of this tongue of unburnt gas.  We were not able to resolve the onset of these transverse detonations.  Photographs obtained at earlier times all showed evidence of transverse detonations, or did not permit to make any conclusion due to lack of resolution.  Their origin remains to be clarified in further studies.  

%%%%%%%%%%%%%%%%%%%%%%%%%%%%%%%%%%%%%%%%%%%%%%%%%%%%%%%%%%%%

\section{Numerical prediction}
In order to gain further insight into the flowfield observed experimentally, we have also performed a numerical reconstruction of the flowfield.  We used the model developed by Radulescu and Maxwell \cite{Radulescu&Maxwell2011}, who also addressed the problem of the interaction of a detonation wave with a cylinder numerically.  The model uses the reactive Euler equations using a single-step Arrhenius reaction model.  The absence of explicit molecular diffusion terms, or turbulence models, restricts the validity of the results to the scales resolved; the grid resolution (i.e. the most resolved grid in the adaptive mesh refinement scheme) is 70 $\mu{m}$.  We modelled the chemistry by calibrating the one-step model against the GRI-3 kinetic mechanism, following the method outlined in \cite{Radulescu&Maxwell2011}.   The parameters used are $Q/RT_0=60.5$, $Ea/RT_0=48.3$ and $\gamma=1.17$, which reproduce the Chapman-Jouguet Mach numbers, reaction sensitivity to temperature changes and the isentropic exponent in the shocked gases.  The effect of varying the initial pressure was monitored by adjusting the pre-exponential constant in order to reproduce the correct physical induction zone length.  

\begin{figure}
\centering
\includegraphics[width=\columnwidth]{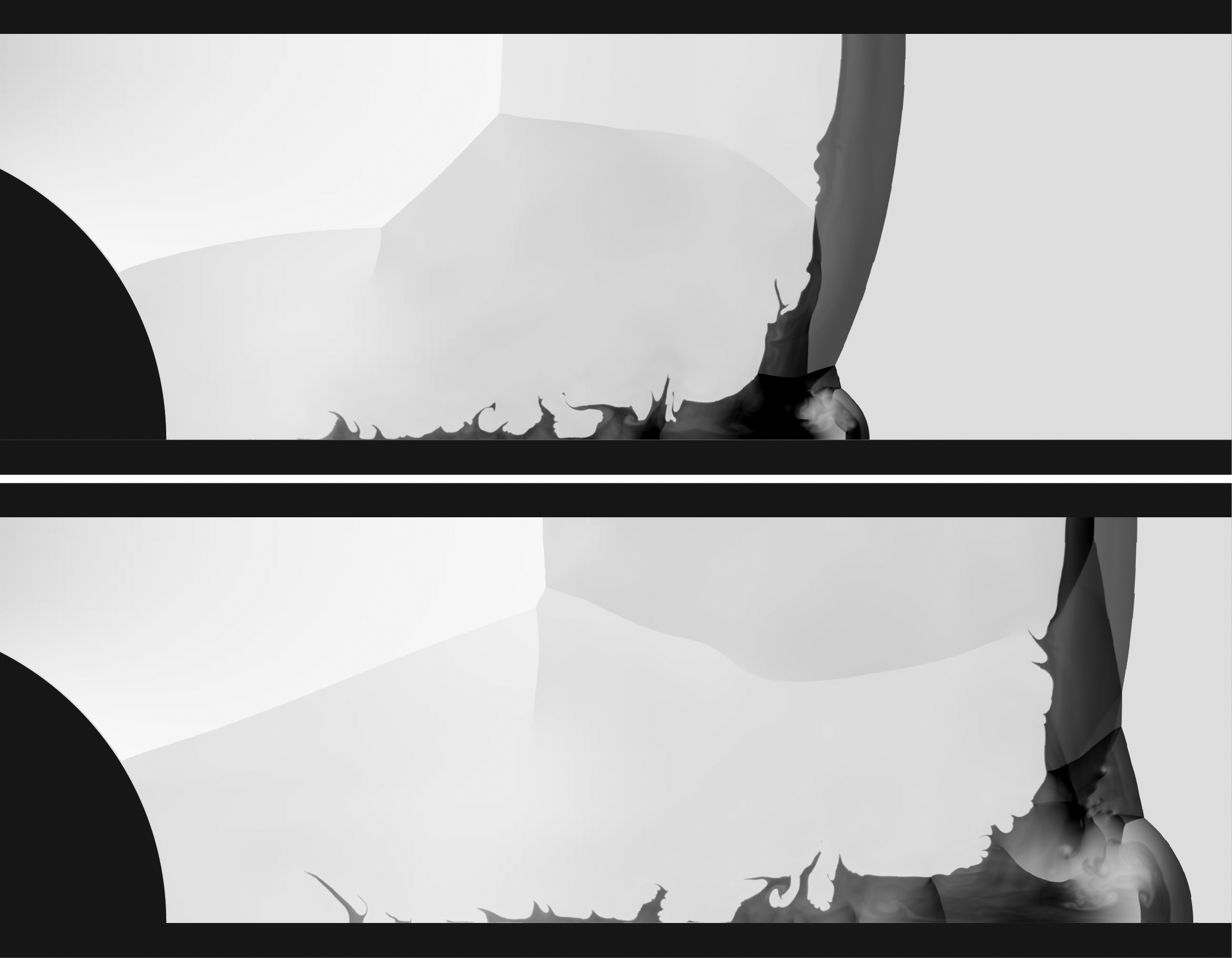}
\caption{Numerical reconstruction of density field at $p_0=$5.5 kPa illustrating the inert shock reflections in the absence of re-ignition.}
\label{fig:numerical5}
\end{figure}

Figure \ref{fig:numerical5} shows the evolution of the density field at an initial pressure of 5.5 kPa.  The simulation reproduces accurately the detonation quenching observed experimentally, shown in Figure \ref{fig:55kpa}.  It also captures the strong wall jet \cite{Collela&Glaz1984} characteristic of strong shock reflections \cite{Mach&Radulescu2011}.  Note however that the jet is much more prominent in these inviscid calculations as compared to the experiment (Fig. \ref{fig:55kpa}).

\begin{figure}
\centering
\includegraphics[width=\columnwidth]{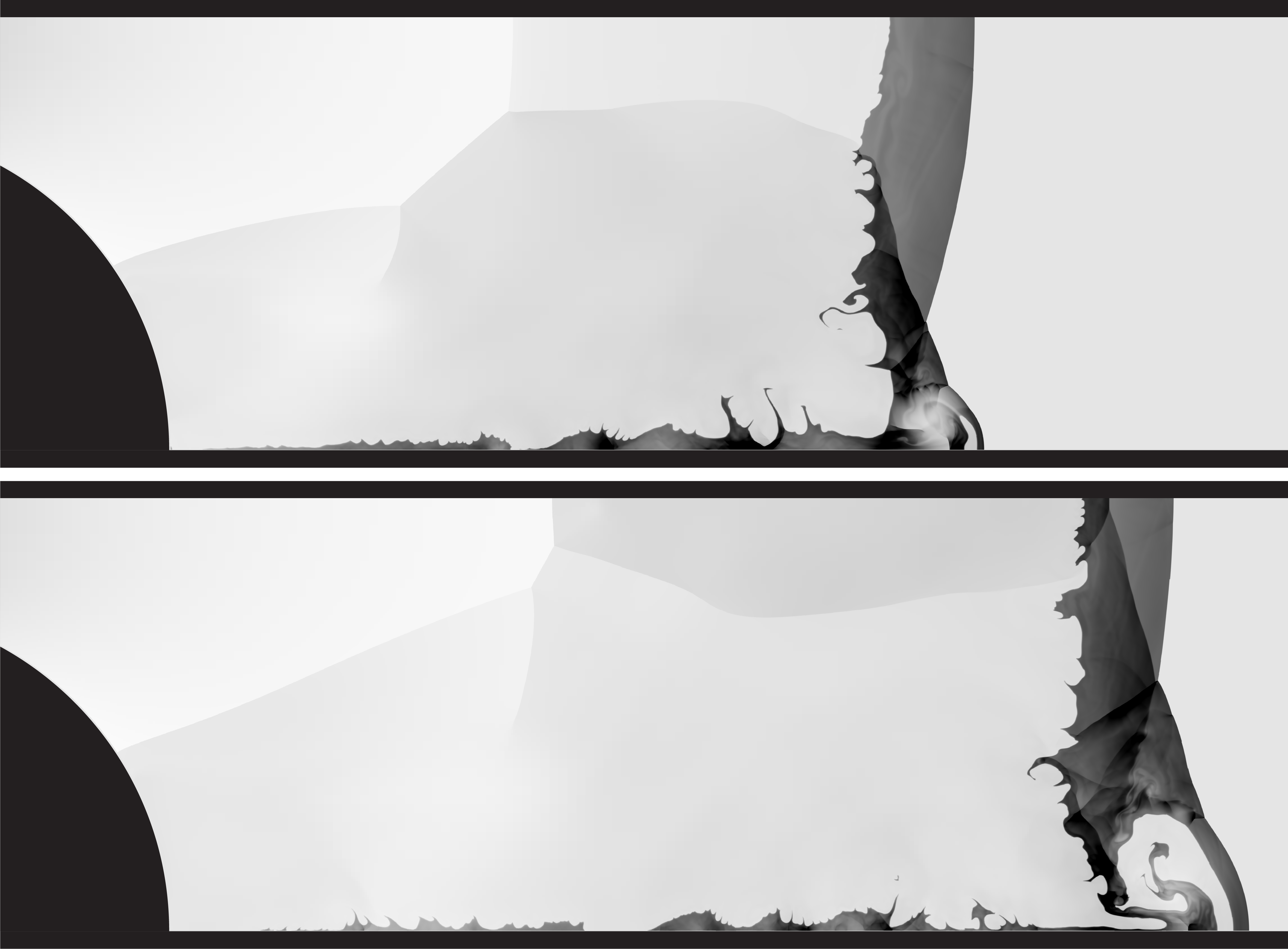}
\caption{Numerical reconstruction of density field at $p_0=$10kPa illustrating the ignition behind the Mach stem and jet entrainment of products.}
\label{fig:numerical10}
\end{figure}

Figure \ref{fig:numerical10} shows the evolution of the density field at an initial pressure of 10 kPa.  At these more reactive conditions, the shock reflection locally ignites the gas behind the Mach stem, consistent with the experimental evidence of Figure \ref{fig:10kpa}.  The two frames in Figure \ref{fig:numerical10} show the localized ignition which occurs behind the Mach shock, and the ensuing entrainment of the gas in the large wall vortex.  We can thus conclude that the first ignition event is indeed due to adiabatic compression.  The strong wall vortex then entrains this gas into a vortex roll.  The evolution of this transient to later times would require the resolution of this complex wave motion with the correct transport terms incorporated with the model.   This is left for future study.
    
%%%%%%%%%%%%%%%%%%%%%%%%%%%%%%%%%%%%%%%%%%%%%%%%%%%%%%%%%%%%  

\section{Conclusions}
The present study clarified the sequence of events leading to the detonation formation when a shock-flame complex undergoes a Mach shock reflection.  Among the mechanisms isolated, the jet formation behind Mach shocks was found to play a very important role in igniting significant amount of gas behind the Mach stem.  When the Mach stem itself is sufficiently strong to permit auto-ignition, a detonation first appears along the Mach stem, while the transverse shock wave remains non-reactive.  The structure of the unburned tongue however indicates local instabilities and hot spot formation, which remain to be elucidated.  The rapid reaction of this gas may be at the origin of the transverse detonations observed in the experiments for more sensitive mixtures. 
%%%%%%%%%%%%%%%%%%%%%%%%%%%%%%%%%%%%%%%%%%%%%%%%%%%%%%%%%%%%
\section*{Acknowledgements}
We wish to thank NSERC for financial support through a Discovery Grant to M.I.R. and the H2CAN Strategic Network of Excellence.  
%% The Appendices part is started with the command \appendix;
%% appendix sections are then done as normal sections
%% \appendix
%% References
%%
%% Following citation commands can be used in the body text:
%% Usage of \cite is as follows:
%%   \cite{key}         ==>>  [#]
%%   \cite[chap. 2]{key} ==>> [#, chap. 2]
%%

%% References with bibTeX database:

\bibliographystyle{elsarticle-num}
\bibliography{references}

%% Authors are advised to submit their bibtex database files. They are
%% requested to list a bibtex style file in the manuscript if they do
%% not want to use elsarticle-num.bst.

%% References without bibTeX database:

% \begin{thebibliography}{00}

%% \bibitem must have the following form:
%%   \bibitem{key}...
%%

% \bibitem{}

% \end{thebibliography}

\end{document}